\documentclass{LMCS}

\usepackage{latexsym,amssymb,bussproofs,hyperref}

\newcommand{\btrue}{{\sf T}}
\newcommand{\bfalse}{{\sf F}}

\newcommand{\bars}[2]{\Phi}
\newcommand{\bark}[2]{\Phi}

\newcommand{\all}[1]{\forall #1\,}

\newcommand{\FV}{{\sf FV}}

\newcommand{\N}{\mathbb{N}}
\newcommand{\B}{\mathbb{B}}

\newcommand{\length}[1]{|#1|}

\newcommand{\modbar}[1]{\Phi}

\newcommand{\overwrite}{@}

\newcommand{\dle}{\sqsubseteq}
\newcommand{\dom}[1]{{\sf dom}(#1)}

\newcommand{\set}[1]{\{#1\}}
\newcommand{\tm}{\subseteq}

\newcommand{\update}[3]{{#1}_{#2}^{#3}}

\newcommand{\nat}{{\sf nat}}
\newcommand{\boole}{{\sf boole}}

\newcommand{\bseq}{[}
\newcommand{\cov}[1]{\overline{#1}} 

\newcommand{\eseq}{]}

\newcommand{\ifthenElse}[3]{{\sf if}\,#1\,{\sf then}\,#2\,{\sf else}\,#3}

\newcommand{\nilseq}{\bseq\eseq}

\newcommand{\snoc}[2]{#1{*}#2}

\newcommand{\const}{\mathcal{C}}
\newcommand{\consto}{\const_{\omega}}
\newcommand{\constr}{\mathcal{CO}}
\newcommand{\co}{{\sf co}}

\newcommand{\succc}[1]{{\sf S}(#1)}
\newcommand{\nilc}{\nilseq}

\newcommand{\conscp}[2]{{\sf cons}(#1,#2)}

\newcommand{\myconv}{\mapsto}
\newcommand{\redone}{\to}
\newcommand{\SN}{{\sf SN}}

\newcommand{\select}[2]{#1_{#2}}

\newcommand{\seval}[1]{[#1]}

\newcommand{\sevalcv}[3]{[#1]^{#2}#3}

\newcommand{\frs}{\mathcal{R}}
\newcommand{\frso}{\frs_{\omega}}
\newcommand{\alphao}{\alpha_{\omega}}

\newcommand{\strat}[2]{#1_{[#2]}}
\newcommand{\omegapprox}{\preceq}
\newcommand{\GT}{T}
\newcommand{\types}{\mathcal{T}}

\newcommand{\lh}{{\sf lh}}
\newcommand{\iget}{{\sf get}}

\newcommand{\ifconst}{{\sf if}}

\newcommand{\append}{+\!\!+}

\newcommand{\listtype}[1]{{\sf list}(#1)}
\newcommand{\TV}{{\sf TV}}

\newcommand{\consd}{{\tt cons}}
\newcommand{\nild}{{\tt nil}}

\newcommand{\abstd}{{\tt abst}}
\newcommand{\abst}{{\sf abst}}
\newcommand{\appd}{{\tt app}}
\newcommand{\constrd}{{\tt constr}}
\newcommand{\maybe}[1]{{\sf Maybe}(#1)}
\newcommand{\just}[1]{{\sf Just}(#1)}
\newcommand{\nothing}{{\sf Nothing}}

\newcommand{\lam}{\Lambda}
\newcommand{\lamcl}{\Lambda_0}
\newcommand{\var}{{\sf Var}}
\newcommand{\cod}{{\tt co}}
\newcommand{\inv}[2]{#1^{{-}1}_{#2}}
\newcommand{\discd}{{\tt case}}
\newcommand{\strict}[1]{{\tt strict}(#1)}
\newcommand{\inve}[1]{#1^{{-}1}}

\newcommand{\arity}{{\sf arity}}
\newcommand{\dummy}{{\tt dummy}}
\newcommand{\typint}[1]{[\![#1]\!]}
\newcommand{\typintv}[2]{[\![#1]\!]#2}

\newcommand{\typcl}{\types_0}
\newcommand{\typsys}[1]{\vdash^{#1}}
\newcommand{\cand}[1]{\widetilde{\mathcal{P}}(#1)}
\newcommand{\MBR}{\mathbf{MBR}}

\newcommand{\pow}[1]{\mathcal{P}(#1)}
\newcommand{\dum}[1]{\widetilde{#1}}
\newcommand{\dB}{\dum{\B}}
\newcommand{\dN}{\dum{\N}}
\newcommand{\dL}[1]{\dum{{\sf list}}(#1)}
\newcommand{\str}[1]{\widehat{#1}}

\def\doi{1 (2:3) 2005}
\lmcsheading%
{\doi}
{14}
{}
{}
{Jan.~25, 2005}
{Sep.~26, 2005}
{}
\begin{document}

\title[Strong normalisation for applied lambda calculi]
      {Strong normalisation for applied lambda calculi}
\author[U.\ Berger]{Ulrich Berger}
\address{Department of Computer Science,
         University of Wales Swansea,
         Singleton Park,
         Swansea,
         SA2 8PP, United Kingdom}
\email{u.berger@swan.ac.uk}




\begin{abstract}
We consider the untyped lambda calculus with constructors and
recursively defined constants. We construct a domain-theoretic model
such that any term not denoting $\bot$ is
strongly normalising provided all its `stratified approximations'
are. From this we derive a general normalisation theorem for applied
typed $\lambda$-calculi: If all constants have a total value, then all
typeable terms are strongly normalising.  We apply this result to
extensions of G\"odel's system $\GT$ and system $F$ extended by
various forms of bar recursion for which strong normalisation was
hitherto unknown.
\end{abstract}

\maketitle

\section{Introduction}
\label{sec-intro}
Extensions of typed $\lambda$-calculi
by data types and recursively defined higher-order functions, 
often called \emph{applied $\lambda$-calculi},
play an important role in logic and computer science.
They are used, for example, to represent formal proofs and to give
computational interpretations of logical and mathematical theories
leading to relative consistency results and estimates of the
strengths of theories in terms of their provably recursive functions
\cite{Godel58,Spector62,Girard71,Troelstra73,Cook93}.
They also form the theoretical backbone of
functional and type-theoretic proof/programming languages
\cite{Constable86,Paulin-Mohring93}.
The most important and often also most difficult problem in the study
of applied $\lambda$-calculi is normalisation, i.e.\ the question
whether every term can be reduced to a normal form with respect to
$\beta$-reduction and the rewrite rules for the 
extended calculus. The best possible result in
this connection is \emph{strong normalisation}, i.e.\ termination of
\emph{every} possible reduction sequence.
A common pattern for proving strong normalisation for an applied
$\lambda$-calculus is to take an existing strong normalisation proof
for the `pure' underlying typed $\lambda$-calculus w.r.t.\
$\beta$-reduction only and adapt it to the applied calculus. A typical
example is the strong normalisation proof for G\"odel's system $T$ of
primitive recursive functionals in simple types \cite{Godel58}
which can be obtained  by adapting
Tait's computability method to primitive recursion
\cite{Troelstra73}. 
Similar methods were applied to prove strong normalisation for the calculus of
constructions extended by inductive types
\cite{Altenkirch93b,BlanquiJouannaudOkada99}.
There are important extensions of system $T$ by stronger recursion
principles, for example Spector's bar recursion~\cite{Spector62},
which also have been treated using adaptations of Tait's computability
method, however at the price of considerable complications. The
difficulty in proving normalisation for bar recursion and similar
recursion schemes lies in the fact that these schemes do not use a recursive
descent along some kind of wellfounded structural ordering, but
rather rely on continuity arguments and the ability to construct, in a
suitable model, infinite sequences by nonconstructive choices.  Since
the computability method amounts to the construction of a syntactic
model (built from strongly normalising terms) which does not satisfy
these requirements, one needs to enrich the model, either by
introducing infinite terms~\cite{Tait71,Vogel76}, or by building the
model from sets of terms instead of single terms
\cite{Bezem85a}. These modifications, which work for Spector's bar
recursion, seem to fail, however, for other recursion principles which also
rely on continuity and choice and which occur
in recent work on computational interpretations of classical choice
and related principles \cite{Berardi98,BergerOliva0x,Berger04}. An
example is modified bar recursion (see (\ref{eq-bar-vogel}) in section
\ref{sec-term}).

In this paper we present a new method for proving strong normalisation
of applied lambda calculi which will allow us to deal with 
modified bar recursion and other forms of recursion.
The method roughly proceeds as follows: Let $\types$ be a
strongly normalising typed $\lambda$-calculus. We assume that
$\types$ is given as a type assignment system for untyped
$\lambda$-terms with constructors, and we require that $\types$ allows
for the (nonrecursive) definition of functions by pattern matching 
on constructors. Let $\frs$ be a
higher-order rewrite system
using pattern matching and (possibly) recursion. Then, to prove strong
normalisation for $\types+\frs$ we
\begin{itemize}
\item[(a)] interpret the untyped terms in a
strict domain model where the constants are interpreted according
to $\frs$,
\item[(b)] show that any term not denoting $\bot$ is strongly normalising,
\item[(c)] prove that all constants are total w.r.t.\ the notion of
totality given by the typing discipline of $\types$
\end{itemize}
While, as we will see, (a) and (b) are always possible, (c) will
depend on the given rewrite system $\frs$.  Now, by (c)~(and the
presumed soundness of the typing discipline w.r.t.\ the model) all
typeable terms are total and hence $\neq\bot$. With (b) it follows
that all typeable terms are strongly normalising.

The advantages of this method lie in its generality
and its manifold modularity aspects:
First, strong normalisation of the underlying typed $\lambda$-calculus
can be proven separately. Second, steps (a) and (b) above are
independent of the typing discipline and can be carried out for any
rewrite system $\frs$. The constructions and proofs involved in (a)
and (b) are elementary (formalisable in primitive recursive
arithmetic). Third, the logical and
mathematical strengths of the typing discipline and the rewrite rules
only enter into step (c). The proof of totality of the constants can
usually be carried out using the intuitive argument why the given
rewrite rules are `semantically sound'. Fourth, the combination of
different rewrite systems for which strong normalisation can be shown 
using our method preserves strong normalisation. This holds
because our method only uses totality of the constants of each rewrite
system separately.
   
Our method appears to be similar to Plotkin's adequacy
proof for PCF~\cite{Plotkin77}, and, in fact, is inspired
by \cite{Plotkin77}. The differences are that Plotkin intertwines
the computability method for the simply typed $\lambda$-calculus with
a semantic approximation argument whereas we keep these arguments
separate. Also, Plotkin deals with full recursion without pattern
matching and shows weak normalisation for closed terms of ground-type
only whereas our recursions are of a restricted form, but we show strong
normalisation for terms of arbitrary types.

As mentioned above our method only applies to rewrite systems based on
a restricted form of pattern matching. These restrictions enforce that
the rewrite rules define the operational and denotational semantics of
the constants in a canonical way. Hence, rules like distributivity or
permutative conversions, which express optimisations rather than
definitions, are not covered by our method. Strong
normalisation results for higher-order rewrite rules of the latter
kind are proven in~\cite{PolSchwichtenberg95} with a different semantic
method based on another notion of strict functionals.

Our paper is organised as follows. Section~\ref{sec-term} introduces
untyped $\lambda$-calculi with constructors and constants together
with higher-order rewrite rules. As our running example we discuss a
few primitive recursive constants and rewrite rules for modified bar
recursion. We give an informal argument why, under the assumption that
higher type functions are continuous, bar recursion is sound. In
section~\ref{sec-model} we define a strict domain-theoretic model for
the calculi introduced previously. We prove that any term not denoting
$\bot$ is strongly normalisable provided all its `stratified
approximations' are, where, roughly speaking, the $n$-th stratified
approximation of a term is obtained by replacing each constant by a
variant for which maximally $n$ unfoldings of the recursion equations
are allowed. The untyped result is used in section~\ref{sec-type} to
prove a strong normalisation theorem for applied $\lambda$-calculi
based on an abstract notion of a strongly normalising and total type
system.  We have kept the notion of type system as general as possible
in order to prepare the ground for future applications of our method
to a variety of type systems.  In section~\ref{sec-applications} we
consider as an example Girard/Reynold's system $F$ of second-order
polymorphic $\lambda$-calculus. We show that extending this system by
higher type primitive recursion and modified bar recursion does not
destroy strong normalisation. We also discuss some other higher-order
rewrite systems our method applies to.

In \cite{Berger0x} we have worked out a special case of our method,
tailored for the simply typed $\lambda$-calculus.  By giving up
generality the definition of the model and the totality proofs then
are slightly simpler. However, we feel that the greater flexibility
gained by the type free approach of this paper is worth paying the
price of technically slightly more involved constructions.

\subsection*{Acknowledgements}
The comments and the constructive criticism by two anonymous referees
contributed significantly to an improved presentation of this work.
%

\section{The type free $\lambda$-calculus with constructors and recursion}
\label{sec-term}
We fix a set $\var$ of variables $x,y\ldots$.  Given a set $\constr$
of \emph{constructors} $\co$
and a set $\const$ of constants $c$, the set of terms $\lam =
\lam(\constr,\const)$ is defined by 
$$ \lam\ni M,N := 
x \mid \co(M_1,\ldots,M_k)\mid c \mid \lambda xM \mid MN$$
where $k$ is the \emph{arity} of $\co$, which is fixed for each constructor.
If $\constr$ is fixed, but $\const$ may vary, then we write
$\lam(\const)$ instead of $\lam(\constr,\const)$.
We let $\FV(M)$ denote the set of free variables of a term $M$.

The operational meaning of the constants $c\in\const$ is given by a 
a set $\frs$ of \emph{rewrite rules} of the form
\begin{eqnarray*}
cP_1\ldots P_n &\myconv& R
\end{eqnarray*}
where $\FV(cP_1\ldots P_n)\supseteq\FV(R)$ and the $P_i$ are
\emph{constructor patterns}, i.e.\ terms built from variables and the
constructors such that no variable occurs twice in the term $c\vec P$.
The number $n$ of arguments $P_i$ is fixed for each constant $c$ and is
called the \emph{arity} of $c$.
We furthermore require
the $P_i$ to be mutually variable disjoint and the left hand sides of
different rules to be non-unifiable. A rule of the form $c\vec P\myconv
R$ is called a \emph{rule for $c$}.

A \emph{conversion}, $M\mapsto N$, is either a $\beta$-conversion, 
$(\lambda x M)N\mapsto_\beta M[N/x]$,
or an instance of
a rewrite rule, i.e.\ $L\sigma\mapsto_\frs R\sigma$ for some
rewrite rule $L\mapsto R\in\frs$ and substitution $\sigma$. 
We write $M\redone_{\frs} N$ if $N$ is obtained from $M$ by replacing one
subterm occurrence of the left hand side of a conversion by
its right hand side.

We call a term $M$ \emph{strongly normalising} with respect to $\frs$,
$\SN_\frs(M)$, if there is no infinite reduction sequence
$M\redone_\frs M_1\redone_\frs\ldots$. Equivalently, the predicate
$\SN_\frs$ is inductively generated by the single rule:
\begin{itemize}
\item[] If $\SN_\frs(N)$ for all $N$ such that $M\redone_\frs N$, then 
$\SN_\frs(M)$.
\end{itemize}
Our goal is to prove strong
normalisation for various classes of terms (which are usually given
by typing disciplines) with respect to various rewrite systems
$\frs$.

As an example consider the constructors $\btrue$, $\bfalse$, 
$0$, $\nilc$, ${\sf S}$, ${\sf cons}$ and the constants 
$\ifconst$, $<$, $\lh$, $\iget$ and $\append$ with the rewrite rules
\begin{eqnarray*}
\ifconst\,\btrue\,x\,y &\ \myconv\ & x\\
\ifconst\,\bfalse\,x\,y &\myconv& y\\[0.4em]
n<0 &\ \myconv\ & \bfalse \\ 
0<\succc{m} &\myconv& \btrue\\
\succc{n}<\succc{m} &\myconv& n<m\\[0.4em]
\lh\,\nilc &\myconv& 0\\
\lh\,\conscp{x}{s} &\myconv& \succc{\lh s}\\[0.4em]
%
%
%
\iget\,\conscp{x}{s}\,0 &\myconv& x\\
\iget\,\conscp{x}{s}\,\succc{n} &\myconv& \iget\,s\,n\\[0.4em]
\nilc\append\ t &\myconv&t\\
\conscp{x}{s}\append\ t &\myconv&\conscp{x}{s\append\ t}
\end{eqnarray*}
When assigning suitable types to these constructors and constants
one obtains a subsystem of G\"odel's system $T$ of primitive recursive
functionals in finite types which is well-known to be strongly
normalising. We are interested in stronger forms of recursion,
for example bar recursion, which we discuss now.
To improve readability will write $\select{M}{k}$ for $\iget M k$,
$\snoc{M}{N}$ for $M\append\ \conscp{N}{\nilc}$ and
$\ifthenElse{M}{N}{K}$ for $\ifconst MNK$. The following form of 
\emph{modified bar recursion} was studied in \cite{Berardi98} and 
\cite{BergerOliva0x}:
\begin{eqnarray*}
\Phi ygs &=& y(\lambda k.\ifthenElse{k<\length{s}}{\select{s}{k}}{gk
(\lambda x.\Phi yg(\snoc{s}{x}))})
\end{eqnarray*}
In order to make sense of this equation one should think of the
variables being typed as follows:
$y\colon(\nat\to\rho)\to\nat$,
$g\colon\nat\to(\rho\to\nat)\to\rho$,
$s\colon \listtype{\rho}$,
where $\rho$ is an arbitrary type.
A functional $\Phi$ satisfying the recursion equation above was used
in \cite{Berardi98} and \cite{BergerOliva0x} to give a realisability
interpretation of the classical (i.e.~negative translated) axiom of
countable dependent choice. Below we give an intuitive argument why,
under the assumption that higher type functions are continuous,
the equation above is sound in the sense that it defines a total
functional (i.e.\ maps total arguments to total values):
\begin{itemize}
\item[]Let $g,y,s$ be total arguments where
$s=[x_0,\ldots,x_{n-1}]$ with total $x_i$, and assume, for
contradiction, that $\Phi ygs$ is undefined. Then the infinite
sequence 
$\lambda k.\ifthenElse{k<\length{s}}
{x_k}
{gk(\lambda x.\Phi yg(\snoc{s}{x}))})$ 
cannot be total, so there must be some $k$ such that $gk(\lambda
x.\Phi yg(\snoc{s}{x}))$ is not total.  Since $g$ is total this
implies that $\lambda x.\Phi yg(\snoc{s}{x})$ is not total, i.e.\
$\Phi yg(\snoc{s}{x_n})$ is not total for some total $x_n$.  Repeating
this argument one arrives at an infinite sequence of total elements $x_k$
such that $\Phi yg[x_0,\ldots,x_{m-1}]$ is undefined for all $m\ge
n$. 
Since all $x_k$ are total,
$y(\lambda k.x_k)$ must be defined.
Furthermore, since $y$ is assumed to be continuous it
will query its argument at numbers $k$ smaller than some fixed number
$m$ only. But then $\Phi yg[x_0,\ldots,x_{m-1}]$ must be defined as
well, which is a contradiction.
\end{itemize}
In the proof of theorem~\ref{theo-sn-bar} we will repeat a slight
variation of this argument in a strict domain-theoretic model in order
to obtain strong normalisation of system $F$ extended by modified bar
recursion.  More precisely, since turning the equation above into a
rewrite rule would clearly not be strongly normalising, we will have
to work with the following minor variation of modified bar recursion. 
We replace the conditional expression on the right hand side by a
call of an auxiliary constant $\Psi$ with an extra (boolean) argument
in order to force evaluation of the test $k<\length{s}$ before the
subterm $\Phi ygh(\snoc{s}{x})$ may be further reduced (Vogel's
trick~\cite{Vogel76}).
\begin{equation}\label{eq-bar-vogel}
\left.\begin{array}{lcl}
\Phi ygs &\myconv& y(\lambda k.\Psi ygsk(k<\length{s}))\\
\Psi ygsk\btrue &\myconv& \select{s}{k}\\
\Psi ygsk\bfalse &\myconv& gk(\lambda x.\Phi yg(\snoc{s}{x}))
\end{array}\right.
\end{equation}
Our results will enable us to easily show that 
all terms that are typeable in system $F$ (under a suitable typing of
the constructors and constants) are strongly normalising
with respect to the rewrite rules above.

\section{A domain-theoretic model for strong normalisation}
\label{sec-model}
By a \emph{domain} we mean a Scott domain, i.e.\ a consistently
complete algebraic domain with an effective base
\cite{Griffor93,AbramskyJung94}.  The least element of a domain $D$ is
denoted $\bot_D$ (or $\bot$, if no confusion is possible) and $\dle_D$
(or $\dle$) denotes the domain ordering. $D\to E$ denotes the domain
of continuous functions from $D$ to $E$. Note that $\bot_{D\to
E}=\lambda a\in D.\bot_E$. If $X$ is an effectively given countable
set, then $X_\bot$ denotes the flat domain $X\cup\set{\bot}$ where all
elements of $X$ are maximal and $\bot(\notin X)$ is the least element.
For any domain $D$ we let $D^X$ be the domain of all functions
from $X$ to $D$, ordered pointwise.
For a continuous function $f\colon D^k\to E$ we define the
\emph{strict version}, $\strict{f}\colon D^k\to E$, by
$\strict{f}(\vec a) := f(\vec a)$ if $\bot\notin\vec a$,
$\strict{f}(\vec a) := \bot$ otherwise. Clearly $\strict{f}$ is again
continuous and $\strict{f}\dle f$. Moreover, $\strict{.}$ is itself a
continuous function on the domain $D^k\to E$.
%
We will also use $\maybe{D} := D + 1 = \set{\just{d}\mid d\in
D}\cup\set{\nothing}\cup\set{\bot}$ where $1$ denotes some 1-element
domain and, in general, $D_1+\ldots+ D_k$
denotes the usual separated sum of domains which has as carrier the
disjoint union of the $D_i$ plus a new bottom element and is ordered
in the expected way.

Given a system $\constr$ of constructors 
we define the domain $D$ by the recursive domain equation
$$ D = \Sigma_{\co\in\constr}D^{\arity(\co)} + (D\to D)$$
The existence of a solution to such an equation is guaranteed by the
fact that in the category of domains and embedding/projection pairs
the separated sum and the continuous function space construction are
continuous functors (co-variant in all arguments) and all such
functors have initial fixed points (up to isomorphism).  By the
definition of $D$ every element of $D\setminus\{\bot\}$ is either of
the form $\cod(\vec a)$ with $\vec a\in D^{\arity(\co)}$, or of the
form $\abstd(f)$ with $f\in D\to D$, and there are continuous
functions $\abstd\colon (D\to D)\to D$, $\appd\colon D^2\to D$ and
$\discd\colon D\to(\constr\cup\set{\abst})_{\bot}$ 
as well as for each constructor $\co$ of arity $k$ continuous
functions $\cod\colon D^k\to D$ and $\inv{\cod}{i}\colon D\to D$
($i=1,\ldots,k$) such that
\begin{itemize}
\item[(i)] $\discd(\cod(\vec a))=\co$, $\discd(\abstd(f))=\abst$,
\item[(ii)] $\inv{\cod}{i}(\cod(\vec a))=a_i$,
\item[(iii)] $\appd(\abstd(f),b) = f(b)$.
\end{itemize}
%
%
If $f\in D^k\to D$, then $\abstd(f)$
stands for $\abstd(\lambda a_1\in D.\ldots\abstd(\lambda a_k\in D.
f(a_1,\ldots,a_k))$. Similarly, 
$\appd(a,b_1,\ldots,b_k)$ abbreviates
$\appd(\ldots\appd(a,b_1)\ldots, b_k)$. 
We define for each term $M\in\lam(\const)$ the \emph{strict
denotational semantics} $\seval{M}\colon D^{\const}\to D^{\var}\to D$
by
\begin{eqnarray*}
\sevalcv{x}{\alpha}{\eta} &=& \eta(x)\\
\sevalcv{c}{\alpha}{\eta} &=& \alpha(c)\\
\sevalcv{\lambda x\,M}{\alpha}{\eta} &=&
\strict{\abstd}(\lambda a\in D.\sevalcv{M}{\alpha}{\update{\eta}{x}{a}})\\
\sevalcv{MN}{\alpha}{\eta} &=&
\strict{\appd}(\sevalcv{M}{\alpha}{\eta}{\sevalcv{N}{\alpha}{\eta}})\\
\sevalcv{\co(M_1,\ldots,M_k)}{\alpha}{\eta} &=&
\strict{\cod}(\sevalcv{M_1}{\alpha}{\eta},\ldots,\sevalcv{M_k}{\alpha}{\eta})
\end{eqnarray*}
The soundness of this definition
rests on the fact that domains and continuous functions
form a cartesian closed category.
\begin{lem}
\label{lem-strict}
Let $M,N\in\lam(\const)$, $\alpha\in D^{\const}$, $\eta\in D^{\var}$, 
$\theta\colon\const\to\const'$, $\alpha'\in D^{\const'}$.
\begin{itemize}
\item[(a)] If $\alpha(c)=\bot_D$ for some constant $c$ in $M$, 
then $\sevalcv{M}{\alpha}=\bot_{D^\var}$.
\item[(b)] $\sevalcv{M[N/x]}{\alpha}{\eta}=
\sevalcv{M}{\alpha}{\update{\eta}{x}{a}}$ 
where $a :=\sevalcv{N}{\alpha}{\eta}$.
\item[(c)] 
$\sevalcv{M\theta}{\alpha'}{}=\sevalcv{M}{\alpha'\circ\theta}{}$
$(M\theta := M[\theta(c)/c\mid c\in\const])$.
\item[(d)] $\seval{(\lambda x M)N}\dle\seval{M[N/x]}$.
\end{itemize}
\end{lem}
\proof 
(a-c) are proved by easy inductions on $M$. (d) follows from (b):
Let $a :=\sevalcv{N}{\alpha}{\eta}$. Then
$\sevalcv{(\lambda x M)N}{\alpha}{\eta}
\dle
\appd(\abstd(\lambda b\in
D.\sevalcv{M}{\alpha}{\update{\eta}{x}{b}}),a) = 
\sevalcv{M}{\alpha}{\update{\eta}{x}{a}} = 
\sevalcv{M[N/x]}{\alpha}{\eta}$.
\qed

Next we define the constant assignment $\alpha_\frs\in D^{\const}$
naturally associated with a rewrite system $\frs$. The values
$\alpha_\frs(c)\in D$ are defined by a simultaneous recursion, i.e.\ 
$\alpha_\frs$ is the least fixed point of a certain continuous operator
on the domain $D^\const$. For a constant $c$ without any rule in
$\frs$ we set $\alpha_\frs(c):=\bot$. The definition of
$\alpha_\frs(c)$ for constants with at least one rule requires
some preparation.
For every vector $\vec P = P_1,\ldots,P_k$ of variable disjoint
constructor patterns we define a continuous `inverse' $\inve{\vec P}\colon
D^k\to\maybe{D^{\var}}$. The definition is by recursion on the number of
constructors in $\vec P$. 
$\inve{\vec x}(\vec a) :=
\just{\update{\bot}{\vec x}{\vec a}}$, where $\update{\bot}{\vec
x}{\vec a}(x_i)=a_i$, and
$$\inve{(\vec x,\co(Q_1,\ldots,Q_n),\vec P)}(\vec a,b,\vec c) :=$$
$$
\left\{\begin{array}{ll}
\inve{(\vec x,Q_1,\ldots,Q_n,\vec P)}
 (\vec a,\inv{\co}{1}(b),\ldots,\inv{\co}{n}(b),\vec c) 
& 
\hbox{if $\discd(b) = \co$}\\
\nothing & \hbox{if $\discd(b) \in (\constr\setminus\set{\co})\cup\{\abstd\}$}\\
\bot & \hbox{if $\discd(b) =\bot$}
\end{array}\right.
$$
\begin{lem}
\label{lem-inv}
\hfill
\begin{itemize}
\item[(a)] If $\inve{\vec P}(\vec a) = \just{\eta}$ and $\vec P$ and 
$\vec Q$ are non-unifiable, then $\inve{\vec Q}(\vec
a)\in\set{\bot,\nothing}$.
\item[(b)] $\inve{\vec P}(\sevalcv{\vec P\sigma}{\alpha}{\eta}) =
\sevalcv{\sigma}{\alpha}{\eta}$ where
$\sevalcv{\sigma}{\alpha}{\eta}(x) :=
\sevalcv{\sigma(x)}{\alpha}{\eta}$. 
\end{itemize}
\end{lem}
\proof 
Easy inductions on the number of constructors in $\vec P$.
\qed 
By lemma~\ref{lem-inv}~(a), the condition that the left hand sides of
different rules for the same constant are non-unifiable implies
that for every constant $c$ of $\frs$-arity $k$ and every $\vec a\in
D^k$ there is at most one rule $c\vec P\mapsto R\in\frs$ such that
$\inve{\vec P}(\vec a) = \just{\eta}$ for some $\eta\in
D^{\var}$. This guarantees the soundness of the following
definition of the values of a constant
$c$ with at least one rule in $\frs$:
$\alpha_\frs(c) := \abstd(f)$ where $f\colon D^k\to D$ is
defined (recursively) by
$$
f(\vec a) =\left\{\begin{array}{ll}
\sevalcv{R}{\alpha_\frs}{\eta}
& 
\hbox{if $c\vec P\mapsto R\in\frs$ and $\inve{\vec P}(\vec a) = \just{\eta}$}\\
\dummy & \hbox{if $\inve{\vec P}(\vec a) = \nothing$ for all
$c\vec P\mapsto R\in\frs$}\\
\bot & \hbox{otherwise}
\end{array}\right.
$$
Here $\dummy$ is some fixed element of $D$ whose value
will be irrelevant in this section.
However, when applying our construction to a particular type system
(Section~\ref{sec-applications}), we will have to choose $\dummy$ in
such a way that it lies in the intersection of all denotations of
types (note that $\dummy$ is independent of the type that might be associated
with the constant $c$).

We set 
$$\sevalcv{M}{\frs} := \sevalcv{M}{\alpha_\frs}$$
\begin{lem}\label{lem-eq-red}
If $M\redone N$, then 
$\sevalcv{M}{\frs}{\eta}\dle\sevalcv{N}{\frs}{\eta}$.
\end{lem}
\proof 
Induction on $M$. 

If $M\mapsto_\beta N$, then we use
lemma~\ref{lem-strict}~(d).

Consider the case of a constant conversion,i.e.\   
$c\vec P\sigma \redone R\sigma$.
Set $\vec a := \sevalcv{\vec P\sigma}{\frs}{\eta}=
\sevalcv{\vec P}{\frs}{\eta'}$ where
$\eta'(x) := \sevalcv{\sigma(x)}{\frs}{\eta}$,
by lemma~\ref{lem-strict}~(b).
By lemma~\ref{lem-inv}~(b), $\inve{\vec P}(\vec a)=\just{\eta'}$.
Therefore,
$\sevalcv{c\vec P\sigma}{\frs}{\eta} 
\dle \appd(\alpha_{\frs}(c),\vec a)
\dle \sevalcv{R}{\frs}{\eta'}
= \sevalcv{R\sigma}{\frs}{\eta}$,
again by lemma~\ref{lem-strict}~(b).

All other cases (conversion of a proper subterm) follow
immediately from the induction hypothesis and the fact that the
functions $\strict{\abstd}$, $\strict{\appd}$ and $\strict{\constrd}$
are monotone.
\qed

The key to our first normalisation result is the approximation of a
given rewrite system by a `stratified' rewrite system, that is a
rewrite system where no recursion occurs. More precisely, let $\frs$
be a rewrite system for a given term system $\lam(\const)$ and define
inductively a constant $c\in\const$ to be \emph{stratified}
(w.r.t. $\frs$) if for every rule $cP_1\ldots P_n \myconv R\in\frs$ the
term $R$ is stratified, i.e.\ contains stratified constants
only. Roughly speaking, stratified rewrite systems allow nothing more
than to define functions by pattern matching and case analysis on
constructors.
$\frs$ is called stratified if all constants
are stratified w.r.t.\ $\frs$.

Let $\frs$ be an arbitrary rewrite system for a system of constants
$\const$. For every constant $c\in\const$ and each $n\in\N$ let $c_n$
be a new constant and set $\consto:=\set{c_n\mid c\in\const,
n\in\N}$. For every term $M\in\lam(\const)$ and $n\in\N$ let
$\strat{M}{n}\in\lam(\consto)$ be the term obtained from $M$ by replacing 
every constant $c$ by $c_n$. We define a stratified rewrite system
for $\consto$ by
\begin{eqnarray*}
\frso &:=& \{c_{n+1}\vec P\myconv \strat{R}{n}\mid c\vec P\myconv R
\in \frs, n\in\N\}
\end{eqnarray*} 
In the following we let $M,N,\ldots$ range over $\lam(\const)$ while
$A,B,\ldots$ range over $\lam(\consto)$.
We write $A\omegapprox M$ if replacing in $A$ each constant $c_n$ by
$c$ yields $M$.
In particular $\strat{M}{n}\omegapprox M$.
\begin{lem}
\label{lem-omegapprox} 
If $A\omegapprox M$ and $A$ contains no constant of the form $c_0$,
then to every $\const$-term $N$ such that $M\redone_\frs N$ there is
a $\const_\omega$-term $A$ such that $A\redone_{\frso} B$ and
$B\omegapprox N$.
\end{lem}
\proof 
Easy induction on $M$. 
\qed 
\begin{lem}\label{lem-approx-sup}
$\sevalcv{M}{\frs}=\bigsqcup_n\sevalcv{\strat{M}{n}}{\frso}$.
\end{lem}
\proof 
By definition, $\alpha_\frs$ is the least fixed point of the
continuous functional $\Gamma_{\frs}\colon D^\const\to D^\const$
defined by $\Gamma_\frs(\alpha)(c) := \bot$ if there is no rule for
$c$ in $\frs$, otherwise $\Gamma_\frs(\alpha)(c) := \abstd(f)$ where
$f\colon D^k\to D$ is defined by
$$
f(\vec a) :=\left\{\begin{array}{ll}
\sevalcv{R}{\alpha}{\eta}
& 
\hbox{if $c\vec P\mapsto R\in\frs$ and $\inve{\vec P}(\vec a) = \just{\eta}$}\\
\dummy & \hbox{if $\inve{\vec P}(\vec a) = \nothing$ for all
$c\vec P\mapsto R\in\frs$}\\
\bot & \hbox{otherwise}
\end{array}\right.
$$
Set $\alpha_n(c) := \alpha_{\frso}(c_n)$. We show
\begin{equation}\label{eq-alpha}
\alpha_n = \Gamma_{\frs}^n(\bot)
\end{equation}
by induction on $n$.
For $n=0$ both sides are $\bot$ (the left hand side $=\bot$ because
there are no rules for constants of the form $c_0$). 
If there is no rule for $c$ in $\frs$, then both sides of
(\ref{eq-alpha}) are again $\bot$. Let now $c$ be a constant with at least
one rule in $\frs$.
By induction hypothesis we have $\Gamma_{\frs}^{n+1}(\bot)(c)=\abstd(f_n)$
where
$$
f_n(\vec a) :=\left\{\begin{array}{ll}
\sevalcv{R}{\alpha_n}{\eta}
& 
\hbox{if $c\vec P\mapsto R\in\frs$ and $\inve{\vec P}(\vec a) = \just{\eta}$}\\
\dummy & \hbox{if $\inve{\vec P}(\vec a) = \nothing$ for all
$c\vec P\mapsto R\in\frs$}\\
\bot & \hbox{otherwise}
\end{array}\right.
$$
(note that the definitions of the functions $f$ and $f_n$ above and the
definition of $f$ after lemma~\ref{lem-inv} differ in the constant environment
under which the term $R$ is evaluated).
Since by lemma~\ref{lem-strict}~(c), 
$\sevalcv{R}{\alpha_n}{\eta} = \sevalcv{\strat{R}{n}}{\frso}{\eta}$
($\alpha_n = \alpha_\frs\circ\theta_n$ where $\theta_n(c) := c_n$)
it follows $\Gamma_{\frs}^{n+1}(\bot)(c) = \alpha_{\frso}(c_n) =
\alpha_n(c)$. 
Now, since $\alpha_\frs$ is the directed supremum of
the $\Gamma_{\frs}^n(\bot)$ it follows, by continuity of the
evaluation function
$\sevalcv{M}{}$, equation~(\ref{eq-alpha}) and lemma~\ref{lem-strict}~(c),
$$\sevalcv{M}{\frs}=\bigsqcup_n\sevalcv{M}{\Gamma_\frs^n(\bot)}=
\bigsqcup_n\sevalcv{M}{\alpha_n}=
\bigsqcup_n\sevalcv{\strat{M}{n}}{\alphao}$$ 
\qed 
\begin{thm}
\label{theo-sn-approx}
If 
$\sevalcv{M}{\frs}\neq\bot$ and all $\strat{M}{n}$ are strongly
normalising w.r.t.\ $\frso$, then $M$ is strongly
normalising w.r.t.\ $\frs$.
\end{thm}
\proof 
Assume $\sevalcv{M}{\frs}\neq\bot$.  By continuity we have
$\sevalcv{M}{\Gamma_{\frs}^n(\bot)}\neq\bot$ for some $n$. By
lemma~\ref{lem-approx-sup} it follows 
$\sevalcv{\strat{M}{n}}{\alphao}{\eta}\neq\bot$ for some $n$.
Since, by assumption, $\strat{M}{n}$ is strongly normalising w.r.t.\
$\frso$ it suffices to show: 
\begin{equation}\label{eq-comm}
\hbox{If $\SN_{\frso}(A)$,
$\sevalcv{A}{\alpha}{}\neq\bot$ and $A\omegapprox M$,
then $\SN_\frs(M)$.}
\end{equation}
We show this by induction on $\SN_{\frso}(A)$.
Assume the hypotheses of~(\ref{eq-comm}).
We need
to show that all one step reducts of $M$ are strongly normalising. So,
assume $M\redone_\frs N$. Since
$\sevalcv{A}{\alpha}{}\neq\bot$ we know, by
lemma~\ref{lem-strict}~(a), that $A$ contains no constant of
the form $c_0$. By lemma~\ref{lem-omegapprox} it follows that
$A\redone_{\frso} B$ with $B\omegapprox N$ for some
$B$. By lemma~\ref{lem-eq-red}
(applied to $\frso$), $\sevalcv{B}{\alpha}{}\neq\bot$, hence we can
apply the induction hypothesis to $B$ and $N$.
\qed

\section{Strong normalisation via typing}
\label{sec-type}
In many cases the premises of theorem~\ref{theo-sn-approx}
can be proven for terms that are typeable in a certain type system.
Our main example will be second-order polymorphism (system $F$), but
any type system that meets a few natural conditions would do as
well. These conditions are isolated in the following notions of an
(abstract) \emph{strongly normalising type system} and the notion of
\emph{totality} of a type system with respect to a \emph{type
interpretation}. The former notion is the requirement that all typeable
terms are strongly normalising for $\beta$-conversion plus
any stratified type-sound rewrite system. For a given type
system this slightly stronger notion of strong normalisation can
usually be obtained by a simple modification of the known proof of strong
normalisation for $\beta$-conversion (for example Girard's candidate method).

Concerning strong normalisation we clearly may restrict our attention
to the set $\lamcl(\const)$ of closed terms in $\lam(\const)$.

A \emph{type system} consists of a set $\types$ of \emph{types} and 
a family of ternary relations
${\vdash}_\const\tm \types^{\const}\times\lamcl(\const)\times\types$
(indexed by constant systems $\const$) 
which is stable under type respecting constant substitutions, that is,
\begin{center}
if $\Delta\vdash_\const M\colon\rho$ and
$\Delta=\Delta'\circ\theta$, then
$\Delta'\vdash_{\const'} M\theta\colon\rho$,
\end{center}
for any $\Delta\in\types^{\const}$,
$M\in\lamcl(\types)$, $\rho\in\types$, $\theta\colon\const\to\const'$ and
$\Delta'\in\types^{\const'}$.

A term $M\in\lamcl(\const)$ is called \emph{typeable} w.r.t.\ 
$\Delta\in\types^\const$ if
$\Delta\vdash_\const M\colon\rho$ for some $\rho\in\types$.
A rewrite system $\frs$ for $\lam(\const)$ is
\emph{type-sound} w.r.t.\ $\Delta\in\types^\const$ if
$\Delta\vdash_\const \lambda\vec x L\colon \rho$ implies
$\Delta\vdash_\const \lambda\vec x R\colon \rho$
for every rule $L\mapsto R\in\frs$ with $\FV(L)=\vec x$ and $\rho\in\types$.
The type system $\types,\vdash$ is called \emph{strongly normalising}
if for any set of constants $\const$, any type assignment
$\Delta\in\types^{\const}$ and any stratified type-sound rewrite
system $\frs$ for $\lam(\const)$ all typeable closed terms are strongly
normalising w.r.t.\ $\frs$.

Next, we consider possible semantics of types in the model $D$ introduced in
the previous section.
Since the value of a closed term does not depend on a
variable assignment, we may, for closed terms $M$, set 
$\sevalcv{M}{\alpha}{}:=\sevalcv{M}{\alpha}{\eta}$ where $\eta\in
D^{\var}$ is arbitrary, for example $\eta:=\bot$. 
A \emph{type interpretation} for $\types$ is a mapping that assigns to
every type $\rho\in\types$ a subset $\typint{\rho}$ of
$D\setminus\set{\bot}$ such that whenever $\Delta\vdash_\const
M\colon\rho$ and $\alpha(c)\in\typint{\Delta(c)}$ for every
$c\in\const$, then $\sevalcv{M}{\alpha}{}\in\typint{\rho}$.

In the following we will call $\alpha\in D^\const$ \emph{total} if 
$\alpha(c)\in\typint{\Delta(c)}$ for all $c\in\const$. 
Similarly we will call $a\in D$ \emph{total} if
$a\in\typint{\rho}$ provided $\rho$ is clear from the context.
\begin{thm}
\label{theo-sn-type}
Let $\types,\vdash$ be a strongly normalising type system and
$\typint{\cdot}\colon\types\to\pow{D\setminus\set{\bot}}$ a type
interpretation.  
Let $\Delta\in\types^{\const}$ be a type assignment and $\frs$ a
type-sound rewrite system such that $\alpha_\frs$ is total. Then
all typeable closed terms are strongly normalising w.r.t.\ $\frs$.
\end{thm}
\proof 
Assume $\Delta\vdash_\const M\colon\rho$. Since $\alpha_\frs$ is total it 
follows that $\sevalcv{M}{\alpha_\frs}{}$ is total, i.e.\ a member of 
$\typint{\rho}$, and hence different from $\bot$.
Define the constant substitutions
$\theta_n\colon\const\to\consto\ (n\in\N)$ and
$\theta'\colon\consto\to\const$ by $\theta_n(c)
:= c_n$ and $\theta'(c_n) := c$. Note that $M\theta_n = \strat{M}{n}$.
Set $\Delta' :=
\Delta\circ\theta'$. Since $\Delta = \Delta'\circ\theta_n$ 
it follows $\Delta'\vdash_{\consto} M\theta_n\colon\rho$
for every $n\in\N$.
Furthermore, the stratified rewrite system $\frso$ is type-sound
w.r.t.\ $\Delta'\in\types^{\consto}$. This can be seen as follows:
Assume $\Delta'\vdash\lambda\vec x L\colon \rho$ for some rule
$L\mapsto R\in\frso$ with $\FV(L)=\vec x$. Since $\Delta'
=\Delta\circ\theta'$ it follows $\Delta\vdash\lambda\vec x
L\theta'\colon\rho$. But $L\theta'\mapsto R\theta'\in\frs$. Hence
$\Delta\vdash\lambda\vec x R\theta'\colon\rho$, since $\frs$ is
type-sound. Since $L\mapsto R\in\frso$ we have $R=R\theta'\theta_n$
for some $n$.  But this implies $\Delta'\vdash\lambda\vec x
R\colon\rho$.  Since we have shown that $\frso$ is type-sound, and
since $M\theta_n$ is typeable and $\types,\vdash$ is assumed to be
strongly normalising, it follows that $M\theta_n$ is strongly
normalising w.r.t.\ $\frso$ for every $n\in\N$. Hence both premises of
theorem~\ref{theo-sn-approx} are satisfied and we may conclude
that $M$ is strongly normalising w.r.t.\ $\frs$.
\qed

\section{Applications}
\label{sec-applications}
As an example of an applied $\lambda$-calculus we consider system $F$
extended by lists and constants with rewrite rules based on pattern
matching.
We consider the same set of constructors 
$\constr = \set{\btrue,\bfalse,0,\nilc,{\sf S},{\sf cons}}$
as in section~\ref{sec-term},
but leave the set of constants $\const$ unspecified for the moment.
Given a set $\TV$ of \emph{type variables} $p,p_1,\ldots$, the set of
(open) \emph{types} is defined by the grammar
$$\rho,\sigma := p \mid \boole \mid\nat\mid 
\listtype{\rho} \mid \rho\to\sigma \mid \all{p}\rho$$
A \emph{context} is a mapping $\Gamma$ from a finite set
$\dom{\Gamma}$ of object variables
to the set of types, written as
$x_1\colon\rho_1,\ldots,x_n\colon\rho_n$.  
The rules for the inductive definition of the typing judgments
$\Delta,\Gamma\vdash_\const M \colon  \rho$, where
$\Delta\in\types^{\const}$, $\Gamma$ is a context, $M\in\lam(\const)$
and $\rho\in\types$, are displayed in 
Figure~\ref{figure-typing-rules}. 
\begin{figure}[ht]
\begin{center}
\AxiomC{}
\UnaryInfC{$\Delta,\Gamma, x\colon \rho \vdash_\const x\colon\rho$}
\DisplayProof
%
\quad
\AxiomC{$c\in\const$}
\UnaryInfC{$\Delta,\Gamma \vdash_\const c\colon\Delta(c)$}
\DisplayProof
\\[1em]
\AxiomC{$\Delta,\Gamma, x\colon \rho \vdash_\const M\colon \sigma$}
\UnaryInfC{$\Delta,\Gamma \vdash_\const \lambda x.M\colon \rho\to\sigma$}
\DisplayProof
%
\quad
\AxiomC{$\Delta,\Gamma \vdash_\const M\colon \rho\to\sigma$}
\AxiomC{$\Delta,\Gamma \vdash_\const N\colon \rho$}
\BinaryInfC{$\Delta,\Gamma \vdash_\const MN\colon \sigma$}
\DisplayProof
\\[1em]
\AxiomC{$\Delta,\Gamma\vdash_\const M\colon \rho$}
\UnaryInfC{$\Delta,\Gamma \vdash_\const M\colon \all{p}\rho$}
\DisplayProof
($p$ not free in $\Gamma$)
\quad
\AxiomC{$\Delta,\Gamma\vdash_\const M\colon \all{p}\rho$}
\UnaryInfC{$\Delta,\Gamma \vdash_\const M\colon \rho[\sigma/p]$}
\DisplayProof
\\[1em]
\AxiomC{}
\UnaryInfC{$\Delta,\Gamma \vdash_\const \btrue\colon\boole$}
\DisplayProof
\quad
\AxiomC{}
\UnaryInfC{$\Delta,\Gamma \vdash_\const \bfalse\colon\boole$}
\DisplayProof
\\[1em]
\AxiomC{}
\UnaryInfC{$\Delta,\Gamma \vdash_\const 0\colon\nat$}
\DisplayProof
\quad
\AxiomC{$\Delta,\Gamma \vdash_\const M\colon\nat$}
\UnaryInfC{$\Delta,\Gamma \vdash_\const \succc{M}\colon\nat$}
\DisplayProof
\\[1em]
\AxiomC{}
\UnaryInfC{$\Delta,\Gamma \vdash_\const \nilc\colon\listtype{\rho}$}
\DisplayProof
%
\quad
\AxiomC{$\Delta,\Gamma \vdash_\const M\colon\rho$}
\AxiomC{$\Delta,\Gamma \vdash_\const N\colon\listtype{\rho}$}
\BinaryInfC{$\Delta,\Gamma \vdash_\const \conscp{M}{N}\colon\listtype{\rho}$}
\DisplayProof
\end{center}
\caption{The typing rules for extended System F}
\label{figure-typing-rules}
\end{figure}

We let $\typcl$ be the set of closed types and define
$$\Delta\typsys{F}_\const M\colon\rho 
\quad\colon{\Leftrightarrow}\quad
\Delta,\emptyset\vdash_\const M\colon\rho\hbox{ and }\rho\in\typcl$$
\begin{prop}
\label{prop-f-sn}
$(\typcl,\typsys{F})$ is a strongly normalising type system.
\end{prop}
\proof 
Clearly $(\typcl,\typsys{F})$ is a type system. In order to see that
it is strongly normalising one easily adapts the proof of
strong normalisation for system $F$ as given, for example in
\cite{Barendregt92} (which is based on Girard's proof \cite{Girard71}
in the $\lambda$-calculus version due to Tait \cite{Tait75}), so as to
accommodate stratified rewrite systems.  We leave details to the
reader. A corresponding proof for simple types is worked out in
detail in \cite{Berger0x}.
\qed 
Our next task is to interpret types in the domain $D$ 
defined in section~\ref{sec-model} (for the
specific set of constructors $\constr$ above).
We define the element $\dummy\in D$, which was left unspecified in
section~\ref{sec-model}, recursively by
$$\dummy = \abstd(\lambda a\in D.\dummy)$$
We set $\dB :=\set{\btrue,\bfalse}\cup\abstd(D)$, where
$\abstd(D):=\set{\abstd(d)\mid d\in D}$, and $\dN :=$
the least subset of $D$ that contains $\set{0}\cup\abst(D)$ and
is closed under the constructor ${\sf S}$. Furthermore, for a subset
$A\tm D$ we set $\dL{A} :=$ the least subset
of $D$ that contains $\set{\nild}\cup\abst(D)$ and contains with $d$
also $\consd(a,d)$ for every $a\in A$. Finally, for $A,B\tm D$ we set
$A\to B := \set{\abstd(f)\in D \mid \all{a\in D}(a\in A\to f(a)\in B)}$.
Set
$$\cand{D} := \set{A\tm D\mid \dummy\in A, \bot\not\in A}$$
Note that $\cand{D}$ contains $\dB,\dN$ and is closed under arbitrary nonempty
intersections and under the operations $A\mapsto\dL{A}$ and 
$(A,B)\mapsto A\to B$ (it is precisely the latter closure condition
together with the requirement that the intersection of all types has to be nonempty
that leads to the somewhat mysterious definition of $\dummy$).

For every type $\rho\in\types$ and type variable assignment
$\tau\colon\TV\to\cand{D}$ we define $\typintv{\rho}{\tau}\in\cand{D}$
by recursion on $\rho$:
\begin{eqnarray*}
\typintv{p}{\tau} &=& \tau(p)\\
\typintv{\boole}{\tau} &=& \dB\\
\typintv{\nat}{\tau} &=& \dN\\
\typintv{\listtype{\rho}}{\tau} &=& 
\dL{\typintv{\rho}{\tau}}\\
\typintv{\rho\to\sigma}{\tau} &=& 
\typintv{\rho}{\tau}\to \typintv{\sigma}{\tau}\\
\typintv{\all{p}\rho}{\tau} &=& 
\bigcap_{A\in\cand{D}}\typint{\rho}{\update{\tau}{p}{A}}
\end{eqnarray*}
\begin{lem}
\label{lem-f-total}
$\typint{.}{}$, restricted to closed types, is a type interpretation
for $(\typcl,\typsys{F})$.
\end{lem}
\proof 
Call $\eta\in D^\var$ total for a type assignment
$\tau\colon\TV\to\cand{D}$ and a context $\Gamma$ if $\eta(x)\in
\typint{\Gamma(\rho)}{\tau}$ for every $x\in\dom{\Gamma}$.  By 
a straightforward induction
on typing derivations one shows that if
$\Delta,\Gamma\vdash_\const M\colon\rho$, then
$\sevalcv{M}{\alpha}{\eta}\in \typint{\rho}{\tau}$ for all $\tau$ and all 
$\alpha$, $\eta$ that are total for $\Delta,\Gamma$.
\qed 
Now let $\const$ consist of the constants 
$\ifconst,<,\lh,\iget,\append,\Phi,\Psi$ and a constant for every 
higher-type primitive recursive functional. Let $\MBR$
be the rewrite system consisting of the 
rewrite rules of section~\ref{sec-term} and the usual rewrite rules
for primitive recursion.
The typing $\Delta$ for the constants is as expected. For example,
writing $c\colon\rho$ for $\Delta(c) = \rho$,
\begin{eqnarray*}
\ifconst &:& \all{p}.\boole\to p\to p\\
< &:& \nat\to\nat\to\boole\\
\lh &:&\all{p}.\listtype{p}\to\nat\\
\iget &:& \all{p}.\listtype{p}\to\nat\to p\\
\append &:& \all{p}.\listtype{p}\to\listtype{p}\to\listtype{p}\\
\Phi &:& \all{p}.((\nat\to p)\to\nat)\to
                 ((p\to\nat)\to p)\to
                 \listtype{p}\to
                    \nat\\
\Psi &:& \all{p}.((\nat\to p)\to\nat)\to
                 ((p\to\nat)\to p)\to
                 \listtype{p}\to
                 \nat\to
                 \boole\to
                    p
\end{eqnarray*}
\begin{lem}
\label{lem-f-type-sound}
$\MBR$ is type-sound for $\typsys{F}$ and $\Delta$.
\end{lem}
\proof 
Immediate, by inspection of the rewrite rules.
\qed 
\begin{thm}
\label{theo-sn-bar}
System $F$ extended by G\"odel primitive recursion and modified bar
recursion is strongly normalising.
\end{thm}
\proof 
By theorem~\ref{theo-sn-type}, proposition~\ref{prop-f-sn} and
lemmas~\ref{lem-f-total} and \ref{lem-f-type-sound} it suffices to
show that $\alpha_{\MBR}$ is total. The proof of totality of
$\alpha_{\MBR}(c)$ for constants
$c\in\set{\ifconst,<,\lh,\iget,\append}$ and, more generally, any
primitive recursive constant is easy and left to the
reader. In the following we will write $c$ instead of
$\alpha_{\MBR}(c)$, $\str{a}(\vec b)$ instead of
$\strict{\appd}(a,\vec b)$ and $\str{\lambda}x.e$ instead of
$\strict{\abstd}(\lambda x.e)$.  
We will also write $[x_0,\ldots,x_{n-1}]$ for
$\consd(x_0,\ldots,\consd(x_{n-1},\nild))$ and call such objects
\emph{proper lists}.
According
to the rewrite rules for $\Phi$ and $\Psi$ we have $\Phi
=\abstd(\varphi)$ and $\Psi = \abstd(\psi)$ where for total arguments
$y,g,s,k$
\begin{eqnarray*}
\varphi(y,g,s) &=& \str{y}(\str{\lambda} k.
                                   \psi(y,g,s,k,k<\lh(s)))\\
\psi(y,g,s,k,\btrue) &=& \iget(g,k)\\
\psi(y,g,s,k,\bfalse) &=& \str{g}(k,\str{\lambda} x.
   \varphi(y,g,s*x)))\\
\psi(y,g,s,k,b) &=&\dummy,\ 
\hbox{if $\discd(b) \in (\constr\setminus\set{\co})\cup\{\abstd\}$}
\end{eqnarray*}
More precisely, since $\Phi$ and $\Psi$ have universal types, we chose
an arbitrary set $A\in\cand{D}$ and take 
$y\in (\dN\to A)\to\dN$,
$g\in (A\to\dN)\to A$,
$s\in \dL{A}$,
and show $\varphi(y,g,s)\in\dN$.  We may assume that $s$ is a proper
list, i.e.\ $s =[x_0,\ldots,x_{n-1}]$ with $x_i\in A$ ($i<n$),
because for other $s\in\dL{A}$ we clearly have
$\varphi(y,g,s)=\dummy\in\dN$.  Assume
$\varphi(y,g,s)\not\in\dN$. Then there must be some $k\in\dN$ such
that $k<\lh(s)=\bfalse$ and $\psi(y,g,s,k,\bfalse)\not\in A$. The
latter can only happen if
$\varphi(y,g,[x_0,\ldots,x_{n-1},x_n])\not\in\dN$ for some $x_n\in
A$.  Repeating this argument one obtains an infinite sequence of
elements $x_n,x_{n+1},\ldots$ such that each $x_i\in A$ and
$\varphi(y,g,[x_0,\ldots,x_m])\not\in\dN$ for any $m\ge n$. Define
a continuous function $f\colon D\to D$
$$f(k) := \left\{\begin{array}{ll}
x_m
& 
\hbox{if $k={\sf S}^m(0)$}\\
\dummy
&
\hbox{if $k={\sf S}^m(\abstd(a))$ for some $m\in\N$ and $a\in D$}\\
\bot & \hbox{otherwise}
\end{array}\right.
$$
Clearly $\abstd(f)\in \dN\to A$. Hence $\str{y}(\abstd(f))\in\dN$. 
Since $\lambda a.\str{y}(a)$ is continuous and $\dN$ is an open subset 
of $D$ there is a finite (compact) approximation $f_0\neq\bot$ of $f$ such
that $\str{y}(\abstd(f_0))=\str{y}(\abstd(f))$. From the first equation
for $\psi$ it follows that there is some
$m\ge n$ such that for $s' := [x_0,\ldots,x_m]$ we have
$\abstd(f_0)\dle \str{\lambda} k.\psi(y,g,s',k,k<\lh(s'))$. 
Therefore $\varphi(y,g,s')=\str{y}(\str{\lambda}
k.\psi(y,g,s',k,k<\lh(s')))\in\dN$
which is a contradiction.  
\qed 
We conclude with a brief discussion of other rewrite systems which
have been used to interpret strong classical analytical principles and
for which our method works as well.  In \cite{Berardi98} the following
recursion was considered, which can be viewed as a more efficient `demand
driven' variant of modified bar recursion. As with modified bar
recursion we use an auxiliary constant replacing the if-then-else
construct used in \cite{Berardi98,Berger04}:
\begin{equation}\label{eq-bbc}
\left.\begin{array}{lcl}
\Phi ygs &\myconv& y(\lambda n.\Psi ygs(n\in\dom{s}))\\
\Psi ygs\btrue &\myconv& f[n]\\
\Psi ygs\bfalse &\myconv& gn(\lambda z.\Phi yg(\snoc{s}{(n,z)}))
\end{array}\right.
\end{equation}
where $y\colon(\nat\to\rho)\to\nat$, $g\colon\nat\to(\rho\to\nat)\to\rho$ and 
$s\colon(\nat\times\rho)^*$ is to
be viewed as the graph of a finite function with $n\in\dom{s}$ and $s[n]$
having the expected meanings.
In \cite{Berger04} it was shown that (\ref{eq-bbc}) can be derived
from the following principle of \emph{open recursion} (again
formulated with an auxiliary constant):
\begin{equation}\label{eq-open}
\left.\begin{array}{lcl}
\Phi y\alpha &\myconv& y\alpha(\lambda n,z.
\Psi y\alpha n z (z{\prec}\alpha n))\\
\Psi y\alpha n z \btrue &\myconv& 
\lambda\gamma.\Phi y (\snoc{\cov{\alpha}n}{z}\overwrite\gamma)\\
\Psi y\alpha n z \bfalse &\myconv& \lambda\gamma.0
\end{array}\right.
\end{equation}
Here $\prec\colon\rho\times\rho\to\boole$ is (the graph of) a wellfounded
relation, $\alpha,\gamma\colon\nat\to\rho$, $\cov{\alpha}n=[\alpha
0,\ldots,\alpha(n-1)]$ and $s\overwrite\gamma = \lambda
k.\ifthenElse{k<\length{s}}{s_k}{\gamma k}$. So, $\Phi$ is recursively
called with arguments of the form
$\snoc{\cov{\alpha}n}{z}\overwrite\gamma$, which are lexicographically
smaller than $\alpha$. Note, however, that the lexicographic ordering
on infinite sequences is not wellfounded. Both recursions,
(\ref{eq-bbc}) and (\ref{eq-open}) have the proof-theoretic strengths
of full second order arithmetic. Their significance rests on the
fact that they can be used to give rather direct realisability
interpretations of strong classical theories: (\ref{eq-bbc}) realises
classical countable choice~\cite{Berardi98}, while (\ref{eq-bbc})
realises \emph{open induction}~\cite{Berger04}, a principle
closely related to Nash-William minimal-bad-sequence
argument~\cite{Nash-Williams63}.

By theorem~\ref{theo-sn-type} and the results of this section, the strong
normalisability of system $F$ plus the recursions above boils down to showing
that the interpretations of (\ref{eq-bbc}) and (\ref{eq-open}) are
total.  The totality of (\ref{eq-open}) can be shown by open induction.
To prove totality of (\ref{eq-bbc}) it is easiest to use the 
reduction to (\ref{eq-open}) given in \cite{Berger04}.

\bibliographystyle{alpha} \bibliography{../database}

\end{document}